\documentclass[preprint,aps,showpacs,nofootinbib]{revtex4}

\usepackage{amssymb}
\usepackage{color}
\usepackage{epsfig}
\definecolor{White}{rgb}{1,1,1}
\definecolor{Red}{rgb}{1,0.1,0}
\definecolor{LightYellow}{rgb}{1,1,.875}
\definecolor{SteelBlue}{rgb}{.273,.508,.703}
\definecolor{navy}{rgb}{0,0,.5}
\definecolor{LightCyan}{rgb}{.875,1,1}
\definecolor{DarkRed}{rgb}{.543,0,0}
\definecolor{HotPink}{rgb}{1,.41,.70}
\definecolor{ForestGreen}{rgb}{.13,.54,.13}
\definecolor{OliveDrab}{rgb}{.42,.55,.14}
\definecolor{MediumBlue}{rgb}{0,0,.80}
\definecolor{RoyalBlue}{rgb}{.25,.41,.88}
\definecolor{DeepSkyBlue}{rgb}{0,.746,1}
\definecolor{Brown}{rgb}{0.545,0.271,0.074}

\def\bea{\begin{eqnarray}}
\def\eea{\end{eqnarray}}
\def\bec{\begin{center}}
\def\ec{\end{center}}

\def\beq{\begin{equation}}
\def\eeq{\end{equation}}

\begin{document}

\begin{flushright}
\today
\end{flushright}

\title{\Large Comment on `A collider signature of the supersymmetric
golden region'}
\author{Won Sang Cho, Yeong Gyun Kim and Chan Beom Park}

\vskip 2cm

\affiliation{\it Department of Physics, KAIST, Daejeon 305-701, Korea}

\begin{abstract}
We show that $Z$ polarization measurement is useful 
for distinguishing the signal decay chain
of the `golden region' \cite{perelstein} 
in the minimal supersymmetric standard model(MSSM) parameter space, 
which involves the stop decay of $\tilde t_2 \rightarrow \tilde t_1 + Z$,
from other decay chains involving the neutralino decay
of $\tilde\chi_2^0 \rightarrow \tilde\chi_1^0 + Z$ in different MSSM scenarios.
\end{abstract}

\maketitle

\begin{figure}[ht]
\vskip 0.4cm
\begin{center}
\epsfig{figure=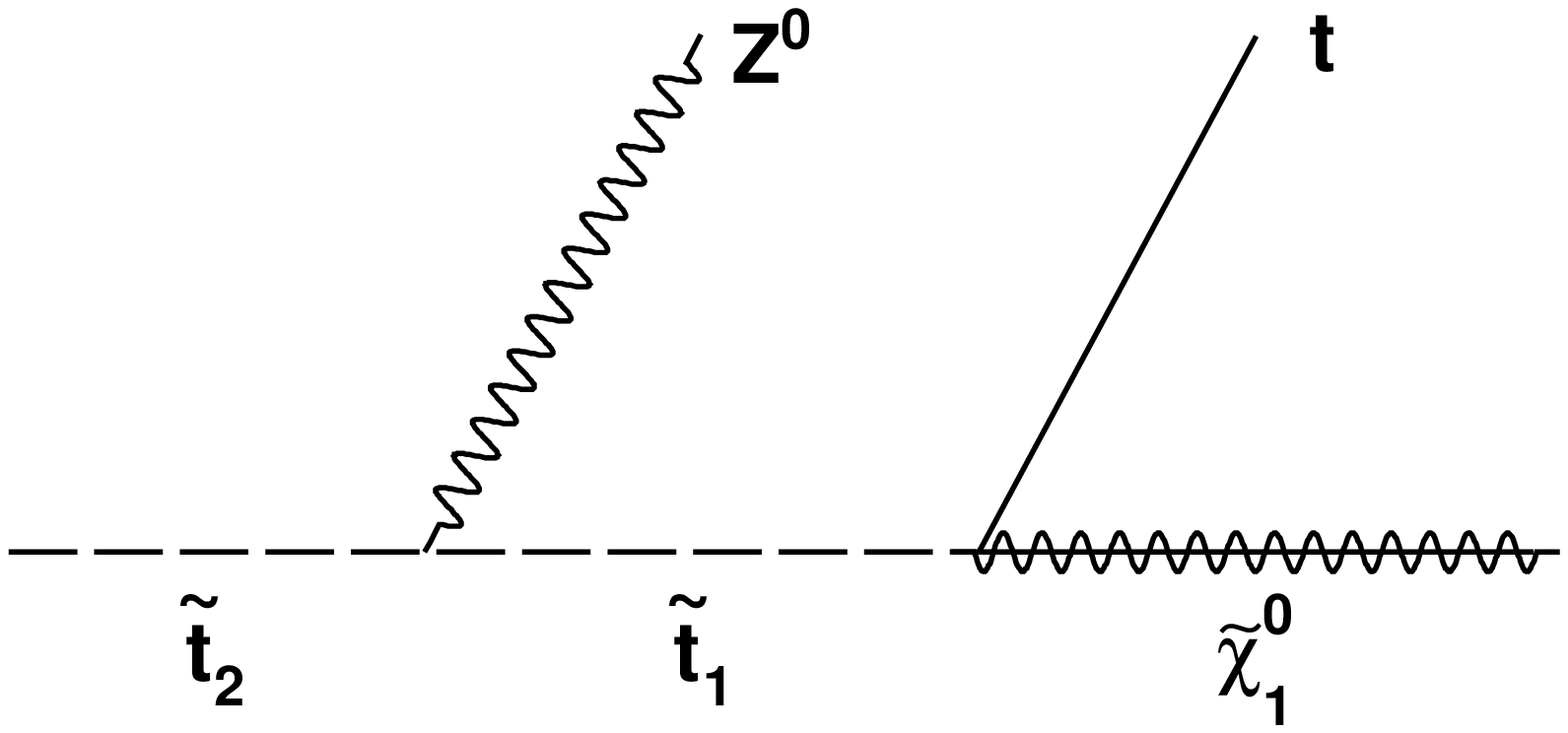,width=7cm,height=5cm}
\epsfig{figure=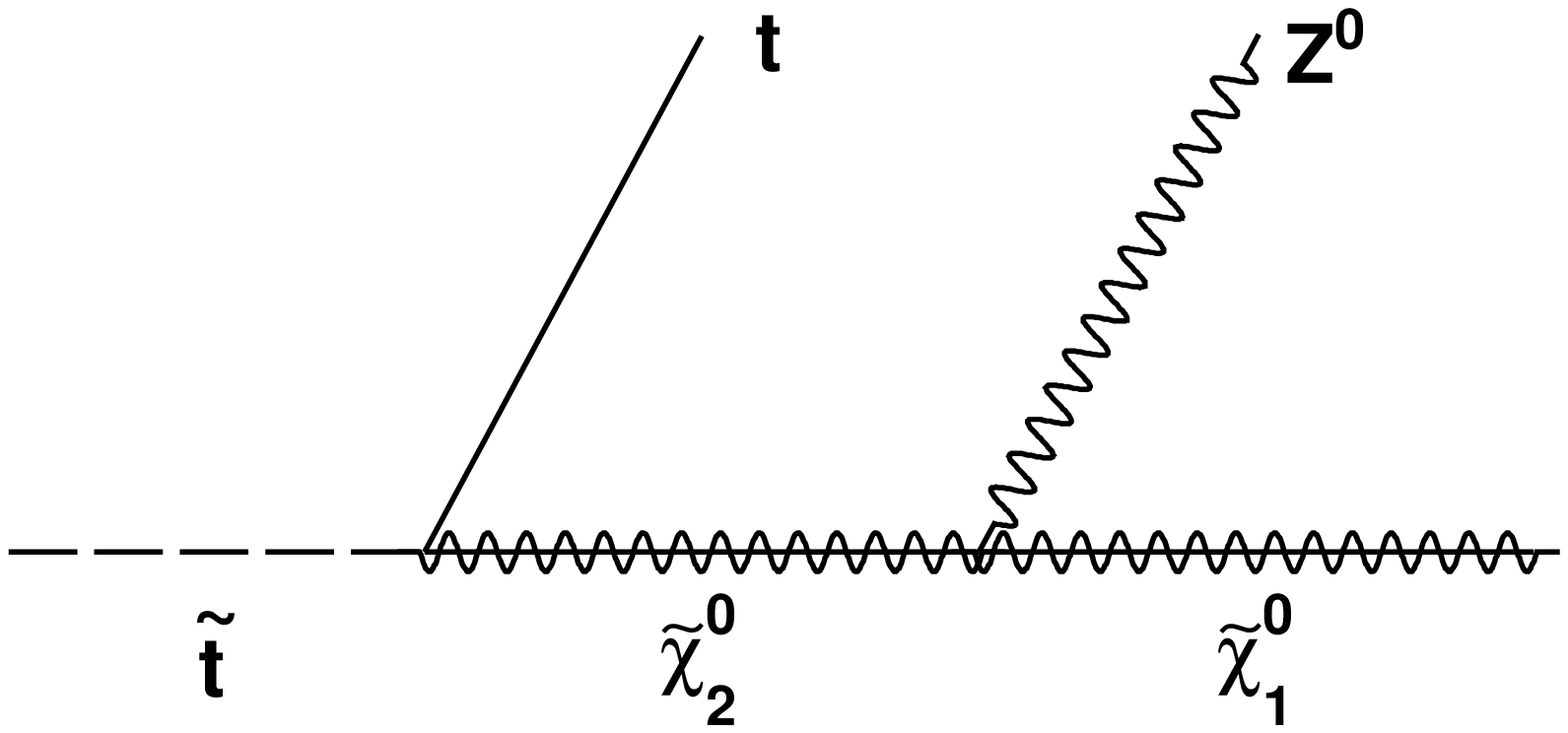,width=7cm,height=5cm}
\end{center}
\caption{\it (a) The decay chain of the golden region
and (b) an alternative decay chain.  }\label{fig:chain}
\end{figure}

Recently, Perelstein and Spethmann investigated
the `golden region' in the MSSM parameter space, where the amount of fine-tuning is minimized
while expermental constraints are satisfied \cite{perelstein}.
The decay of a heavier stop into a lighter stop and a $Z$ boson, 
$\tilde t_2 \rightarrow \tilde t_1 Z$,
is kinematically allowed through the golden region. 
For instance, a benchmark point of the golden region in Ref. \cite{perelstein}
provides stop masses, $m_{\tilde t_2}=700$ GeV, $m_{\tilde t_1}=400$ GeV 
with branching ratio $B({\tilde t_2} \rightarrow {\tilde t_1} Z) = 31 \%$.
Then, the cascade decay in Fig.\ref{fig:chain} (a) provides 
a characteristic signature of the golden region scenario at the LHC.
However, the interpretation of the golden region signature can be obscured by 
other types of cascade decays such as the one in Fig.\ref{fig:chain} (b) 
involving the neutralino decay $\tilde\chi_2^0 \rightarrow \tilde\chi_1^0 Z$,
which is realized in different MSSM scenarios but produces
the same final state particles as the chain in Fig.\ref{fig:chain} (a). 
In this note, we show that one way of distinguishing 
the two confusing decay chains in Fig.\ref{fig:chain}, is to investigate
$Z$ boson polarization in the decay chains. 

For the stop decay $\tilde{t}_2 (p) \rightarrow \tilde{t}_1 (q) + Z (k,\lambda)$,
where $p,q$ and $k$ denote 4-momenta of the particles, and $\lambda=\pm 1,0$ indicate
the helicities of $Z$ boson, 
the matrix element is given by
\begin{eqnarray}
{\cal M} \propto (p+q)^\mu \epsilon^*_\mu (k,\lambda).
\end{eqnarray}
For transverse polarizaion vectors $\epsilon (k,\lambda)$ of $Z$ boson,
we have $p\cdot\epsilon(k,\pm 1)=q\cdot\epsilon(k,\pm 1)=0$ in 
the rest frame of the decaying stop $\tilde t_2$, so that
transverse $Z$ polarization does not appear but only longitudinally 
polarized $Z$ boson is produced from the stop decay in the rest frame. 
In the laboratory frame where $\tilde t_2$ has non-zero spatial momentum,
transverse $Z$ polarization is developed through 
the so-called Wigner rotation \cite{wigner} 
even though it is absent in the stop rest frame.
However, because of large mass hierarchy between the decaying stop and $Z$ boson
for the golden region scenario, the Wigner rotation effect
is so negligible that the $Z$ boson from the stop decay in the laboratory frame
is still dominantly longitudinal. 

On the other hand, both of transverse and longitudinal $Z$ polarizations are available
in $\tilde\chi_2^0 \rightarrow \tilde\chi_1^0 + Z (\lambda)$ decay processes.
In the rest frame of decaying neutralino, 
the relevant decay rates are given by \cite{choikim1}
\begin{eqnarray}
\Gamma[\tilde\chi_2^0 &\rightarrow& \tilde\chi_1^0 + Z(\pm 1)]~ \propto~
m_2^2+m_1^2-m_Z^2- 2 m_2 m_1 {\cal A_N}, \\
\Gamma[\tilde\chi_2^0 &\rightarrow& \tilde\chi_1^0 + Z(0)]~ \propto~
m_2^2+m_1^2-m_Z^2-2 m_2 m_1 {\cal A_N}+ {\lambda_Z \over m_Z^2},
\end{eqnarray}
where $m_1$ and $m_2$ are the masses of $\tilde\chi_1^0$ and $\tilde\chi_2^0$ respectively,
$\lambda_Z \equiv [(m_2+m_1)^2-m_Z^2] [(m_2-m_1)^2-m_Z^2]$ and 
${\cal A_N} \equiv (|V|^2 -|A|^2)/(|V|^2+|A|^2)$.
And the vector and axial vector couplings $V$ and $A$ 
of the $Z$ boson to the neutralino currents
are expressed in terms of the neutralino mixing matrix elements \cite{choi}.
Although the relative production rates of longitudinal 
and transverse $Z$ polarizations in the 
$\tilde\chi_2^0 \rightarrow \tilde\chi_1^0 + Z$ decay depend on 
the neutralino mass spectrum, the $\tilde\chi_2^0 \tilde\chi_1^0 Z$ couplings 
and the $\tilde\chi_2^0$ momentum, 
they are expected to be similar in magnitude
unless $m_2$ is too large compared to $m_Z$.

\begin{figure}[ht]
\vskip 0.4cm
\begin{center}
\epsfig{figure=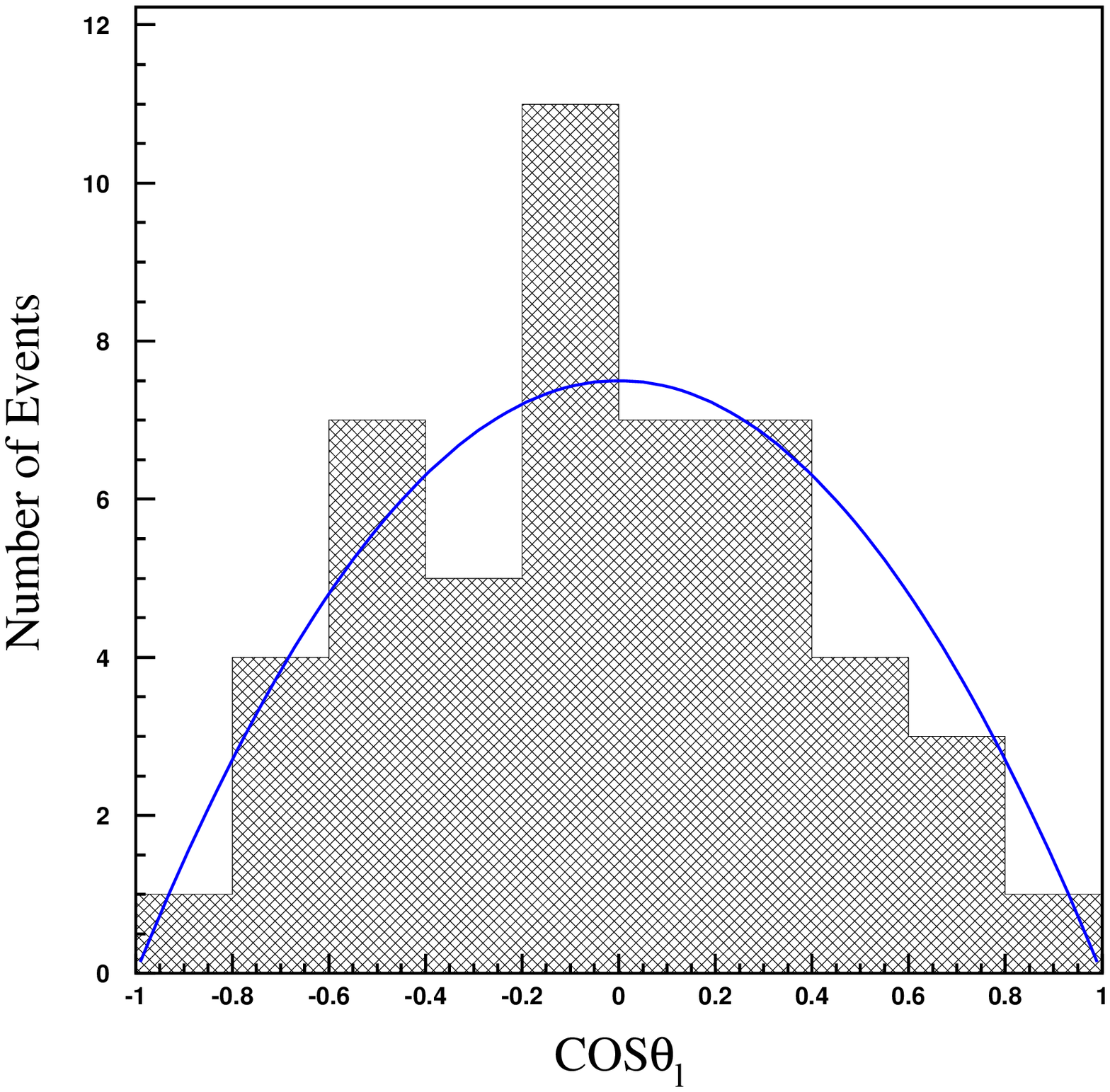,width=7cm,height=7cm}
\epsfig{figure=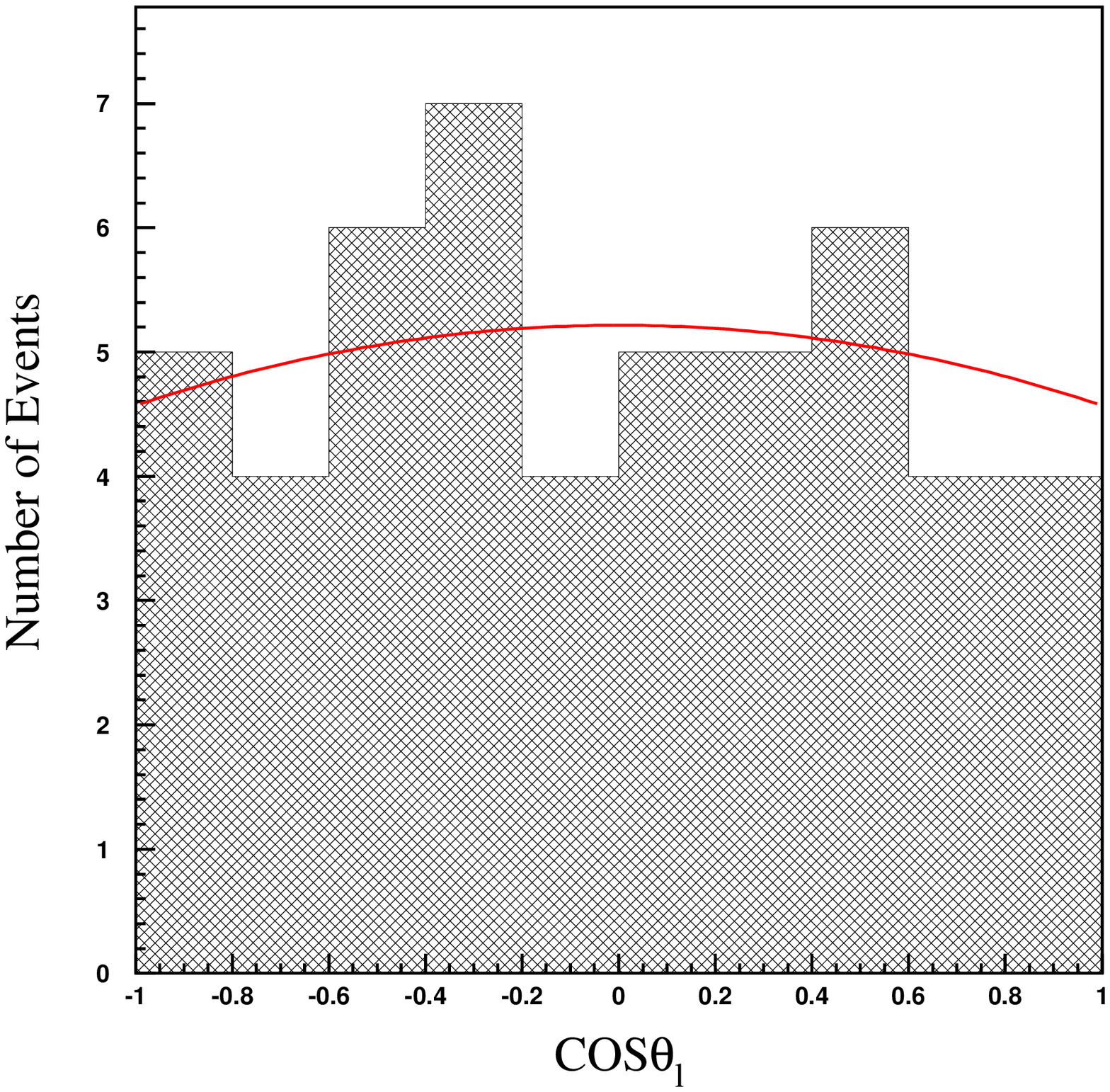,width=7cm,height=7cm}
\end{center}
\caption{\it The lepton angular distributions in the $Z$ rest frame
with respect to $Z$ polarization direction for
(a) the stop decay $\tilde t_2 \rightarrow \tilde t_1 + Z$ and 
(b) the neutralino decay $\tilde\chi_2^0 \rightarrow \tilde\chi_1^0 + Z$, 
respectively. 50 Monte Carlo events are used for the distributions. }\label{fig:zpol}
\end{figure}

The $Z$ polarization can be reconstructed through 
lepton angular distribution in leptonic $Z$ decays $Z \rightarrow l^+ l^-$ \cite{choikim1}. 
Fig.\ref{fig:zpol} (a) shows the leptonic angular distribution of $Z \rightarrow l^+ l^-$ decay 
in the $Z$ rest frame 
with respect to the $Z$ polarization direction, where $Z$ bosons come from
$\tilde t_2 \rightarrow \tilde t_1 + Z$ decays in the golden region scenario. 
On the figure, the histogram corresponds to the result of a parton-level 
Monte Carlo simulation with HERWIG event generator \cite{herwig}, 
while the solid line to the theoretical expectation which ignore the Wigner rotation effect.
Here, 50 Monte Carlo events are used for the histogram.
The number of events are expected for the benchmark point in Ref. \cite{perelstein}
with 300 ${\rm fb}^{-1}$ luminosity if suitable event selection cuts are imposed. 
(See Ref.\cite{perelstein} for the details.) 
However notice that, in this work, detector responce and event selection cuts are not included. 

On the other hand, the corresponding lepton angular distribution 
for the case where $Z$ is produced from neutralino decay of 
$\tilde\chi_2^0 \rightarrow \tilde\chi_1^0 + Z$, is shown in Fig.\ref{fig:zpol} (b).
Here, as a specific numerical example, we set the neutralino masses 
as $m_{\tilde\chi_2^0} = 221.6$ GeV and $m_{\tilde\chi_1^0}=118.5$ GeV
and assume that the neutralino mixing matrix is real. 
Again, on the figure, the histogram corresponds to 50 Monte Carlo events and
the solid line to theoretical expectation without the Wigner roation effect. 

The two decay chains in Fig.\ref{fig:chain} exhibit very
distinctive lepton angular distributions in our example 
so that they can be clearly distinguished experimentally, 
even though more careful studies 
including detector responce, event selection cuts and the SM background
would need to be performed.

In summary, we showed that $Z$ polarization is useful to discriminate the signal
decay chain of the golden region scenario which involves the stop decay
$\tilde t_2 \rightarrow \tilde t_1 Z$,  
from other decay chains including the neutralino decay 
$\tilde\chi_2^0 \rightarrow \tilde\chi_1^0 Z$ in different MSSM scenarios. 

\begin{acknowledgments}
We thank S.Y. Choi for his valuable comments on the manuscript.
This work was supported by the Korea Research Foundation Grant
funded by the Korean Government (MOEHRD, Basic Research Promotion
Fund) (KRF-2005-210-C000006), the Center for High Energy Physics of
Kyungpook National University, and the BK21 program of Ministry of
Education.

\end{acknowledgments}

\end{document}